\documentclass[12pt]{iopart}
\usepackage{iopams}
\usepackage{graphicx}
\usepackage{subfigure}
\usepackage{bm}
\begin{document}

\title{Laser pulse amplification and dispersion compensation in an effectively extended optical cavity containing Bose-Einstein condensates}

\author{D Tarhan$^1$, A Sennaroglu$^2$, \"O E M\"ustecapl{\i}o\u{g}lu$^2$}

\address{$^1$Harran University, Department of Physics, 63300, \c{S}anl\i{}urfa, Turkey}
\address{$^2$Ko\c{c} University, Department of Physics, 34450, Sar{\i}yer, Istanbul, Turkey}

\ead{omustecap@ku.edu.tr}

\begin{abstract}

We review and critically evaluate our proposal of a pulse amplification scheme
based on two Bose-Einstein condensates inside the resonator of a mode-locked
laser. Two condensates are used for compensating the group velocity dispersion.
Ultraslow light propagation through the condensate leads to a
considerable increase in the cavity round-trip delay time, lowers the
effective repetition rate of the laser, and hence scales up the
output pulse energy. It has been recently argued that atom-atom interactions would make
our proposal even more efficient. However, neither in our original proposal nor in the case of interactions,
limitations due to heating of the condensates by optical energy absorption were taken into account.
Our results show that there is a critical time of operation, $~0.3$ ms, for the optimal amplification factor,
which is in the order of $\sim 10^2$ at effective condensate lengths in the order of $\sim 50$ $\mu$m.
The bandwidth limitation of the amplifier on the minimum temporal width of the pulse that can
be amplified with this technique is also discussed.

\end{abstract}

%
%
\submitto{\jpb}
\maketitle

\section{Introduction}

High peak-power laser pulses are sought in diverse scientific and
technological applications such as biomedical imaging \cite{tearney_vivo_1997},
ultrafast spectroscopy \cite{drescher_time-resolved_2002}, and high harmonic generation
\cite{spielmann_generation_1997}. Ultrashort laser pulses in the picosecond to
femtosecond  range are routinely produced by the technique of mode
locking, where the axial modes of the cavity are locked in phase to
produce a train of pulses with a repetition rate $f_{\rm rep}=c/2L$,
where $c$ is the speed of light and $L$ is the effective optical
path length of the resonator \cite{haus_mode-locking_2000}. Pulse energies directly
generated from such lasers are typically in the nJ range and
amplification schemes are necessary in order to scale up the peak
intensities to levels where nonlinear interactions can be observed.
Recently, a simple amplification method was demonstrated, based on
the extension of the standard mode-locked resonator length with a
compact multipass cavity \cite{cho_low-repetition-rate_1999}. If the added insertion loss of
the extended cavity is negligible, the average output power of the
laser remains nearly the same and extension of the cavity length
scales up the energy per pulse by lowering the repetition rate.
While this technique has been widely used to amplify femtosecond
pulses up to  $\mu$J energies \cite{naumov_approaching_2005}, there is an ever growing
need to find amplification methods that will facilitate the
construction of even more compact, high-intensity lasers.

About five years ago, we proposed a compact  laser pulse amplification scheme
using two Bose-Einstein condensates (BECs) introduced in the resonator of a
mode-locked laser \cite{sennaroglu_laser_2007}. The basic idea was to utilize the
electromagnetically induced transparency (EIT)
scheme \cite{harris_electromagnetically_1997,marangos_electromagnetically_1998,fleischhauer_electromagnetically_2005,lukin_resonant_2000}
in a cavity \cite{wang_cavity-linewidth_2000} to make the pulse
ultraslow \cite{hau_light_1999,liu_observation_2001} such that the effective pulse repetition
rate could be drastically reduced due to the increase of the cavity round-trip time,
provided that the average output power of the laser remains nearly the same after the introduction
of the condensates, this increase in round-trip time then scales up the output energy.

The promise of laser pulse amplification by atomic condensates
is further studied by taking into account atom-atom interactions in f-deformed condensate formalism very recently \cite{haghshenasfard_controlling_2012}.
It is found that interactions make the amplification effect stronger. Moreover, in addition to subluminal propagation, possibility of superluminal speeds and high
degree of control over the repetition rate are also shown to be possible. On the other hand, neither of these works consider the effects of the
heating of the condensates due to optical pulse absorption. Present work
is a critical evaluation of the feasibility of the proposal by taking into account
the temperature increase of the condensates during pulse propagation.

In the design of a practical BEC pulse amplifier, certain
effects need to be further considered. First, because
the BEC would introduce an excessive amount of dispersion to short
pulses \cite{tarhan_dispersive_2006}, undesirable pulse broadening results. To
overcome this limitation, we offer a method for dispersion
compensation by using a second intracavity BEC with the proper
choice of energy levels to provide the opposite sign of
dispersion. Second BEC can be in fact prepared identical with the
first, and a magnetic field can be utilized to shift the energy
levels of the second BEC to match the desired detuning of the
probe pulse. Second, the transparency bandwidth of the EIT process
puts a limitation on the shortest temporal pulsewidth that can be amplified
in the resonator. Resulting bandwidth limitation of the amplifier
is further discussed. Our results show that, pulse energy amplification factors
of $\sim 10^2$ are possible by using a $^{23}$Na BEC with effective
lengths of $\sim 10-100\,\mu$m. Finally, heating
of the condensate, due to small but nonzero absorption in the EIT scheme,
introduces an operational time beyond which the condensate turns into
usual Maxwell-Boltzmann gas.
Our calculations reveal a critical time, $~0.3$ ms, at which the amplification
factor is optimum.

The significance of proposed method of effectively increasing the length of the cavity against the straightforward solution of extended cavities should be clarified. In fact it is of high demand to have compact laser cavities in practical laser systems. In applications like optical data storage, filtering or other optical logic and signal processing, having long optical paths, for long optical delay, while maintaining cavity volume small brings tremendous practical advantages. The search for such compact cavities goes back to mid 1960s. Herriott et al. introduced the method of folding long optical paths for compact cavities \cite{herriott1965}. Around early 2000, multi-pass cavities were developed by Cho et al \cite{cho_low-repetition-rate_1999}. This is now an active modern research field (for a review see Ref. \cite{sennaroglu_compact_2004}.). In ultraslow light scheme by electromagnetically induced transparency, optical pulses slow down to speeds about few meters/second. Thus the effective optical path is indeed too long for considering an equivalent extended cavity. In addition to the typical applications of compact laser cavities in accurate optical loss measurements, stimulated Raman scattering, long-path absorption spectroscopy, high-speed path-length scanning,
present system of ultraslow light has another crucial application area of coherent optical memory. With modest amplification power and dispersion compensation, proposed scheme can also be used for that purpose in the context of correcting pulse shape errors in the stored optical information.

This paper is organized as follows. Method of generating ultraslow light via EIT scheme is introduced briefly in Sec. \ref{sec:EITultraslowLight}. We describe our proposed scheme in Sec. \ref{sec:ourScheme}.
The results of dispersion compensation, heating rate, amplification factor and spectral bandwitdth calculations are presented in the corresponding subsections of Sec. \ref{sec:results}. We conclude in Sec. \ref{sec:conclusion}.
\section{Ultraslow light by EIT scheme}\label{sec:EITultraslowLight}

In the EIT configuration, a condition of weak probe is usually
assumed, such that, a strong drive field with Rabi frequency
$\Omega_c$ and the circulating resonator field of frequency
$\Omega_p$ satisfy $\Omega_c\gg \Omega_p$ \cite{scully_quantum_1997}.
A conventional configuration is the case where the probe field is pulsed while
the drive field is continuous wave (CW) \cite{fleischhauer_electromagnetically_2005}.
Studies and
experiments in the strong probe regime indicate deterioration of
EIT and enhanced absorption of the probe pulse \cite{wielandy_investigation_1998}.
Requirements for initiation of EIT is formulated in terms of Rabi
frequencies \cite{harris_preparation_1995}. We are assuming that Rabi frequency of the amplified pulsed
signal remains sufficiently smaller  than the Rabi frequency of the cw drive so that the
EIT conditions are maintained during the operation.
Our results reveal a maximum of amplification factor and the
strength of the cw control field can be chosen accordingly.Though
we did not specifically investigate it here, our scheme may also
be used for compensating enhanced absorption in the strong probe
EIT experiments. This would be particularly advantageous to
examine interactions of probe and coupling fields \cite{kasapi_electromagnetically_1995},
adiabatons \cite{grobe_formation_1994},  as well as to facilitate nonlinear
processes demanded for EIT
applications \cite{li_enhancement_1996,ham_enhanced_1997,harris_optical_1997,zhang_nonlinear_1997}.

Under the weak probe condition, susceptibility $\chi$ for the probe
transition can be calculated as a linear response as most of the
atoms remain in the lowest state. Assuming local density
approximation, neglecting local field, multiple scattering and
quantum corrections and employing steady state analysis, $\chi $ is
found to be \cite{scully_quantum_1997}
\begin{eqnarray}
\chi  = \frac{{\rho \left| {\mu _{31} } \right|^2 }}{{\varepsilon _0
\hbar }} \frac{{i(-i\Delta  + \Gamma _2 /2)}}{{{(\Gamma _2 /2 -
i\Delta ) (\Gamma _3 /2 - i\Delta ) + \Omega _C ^2 /4} }},
\end{eqnarray}
where $\rho$ is the atomic density of the condensate,
$\Delta=\omega_{p}-\omega_{31}$ is the frequency detuning of
the probe field with frequency $\omega_p$ from the resonant
electronic transition $\omega_{31}$, $\Gamma_2$ and
$\Gamma_3$ respectively denote the dephasing rates of the atomic coherences of
the lower states, $\mu _{31}  = 3\varepsilon _0 \hbar \lambda
_{31}^2 \gamma /8\pi ^2 $ is the dipole
matrix element between upper state $|3\rangle$ and lower state
$|1\rangle$, involved in the probe transition with $\lambda_{31}$ being
the resonant wavelength of the
probe transition and $\gamma$ being the radiation decay
rate between $|3\rangle$ and $|1\rangle$.  For the cold gases considered
in this paper and assuming co-propagating laser beams, Doppler
shift in the detuning is neglected.

At the probe resonance, imaginary part of
$\chi$ becomes negligible and results in turning an optically
opaque medium transparent. Furthermore, EIT can be used to achieve
ultra-slow light velocities, owing to the steep dispersion of the
EIT susceptibility $\chi$ about the probe resonance \cite{liu_observation_2001}.
\section{Our proposed scheme of pulse amplifier}\label{sec:ourScheme}

\begin{figure}
\centering{\vspace{0.5cm}}
\includegraphics[width=3.25in]{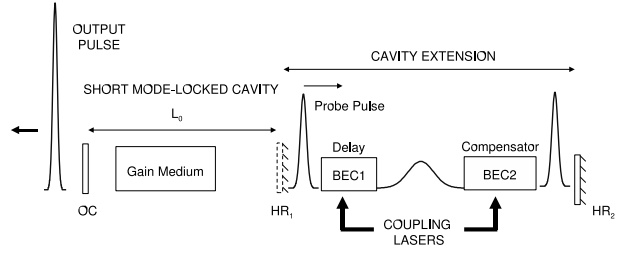}
\caption{Schematic of a mode-locked laser cavity containing two
Bose-Einstein condensates to reduce the pulse repetition rate and
to compensate for dispersion. BEC1 introduces delay to lower the
repetition rate and scales up the pulse energy.} \label{fig1}
\end{figure}

A schematic of the proposed BEC pulse amplifier is shown in Fig.
\ref{fig1}. The short cavity, initially extending from the output
coupler (OC) up to the end high reflector (HR$_1$), contains a
gain medium and is first passively mode-locked by using the
technique of Kerr lens mode locking or a saturable absorber
\cite{haus_mode-locking_2000}.
The cavity is then extended by removing HR$_1$ and two different
BECs are introduced inside the resonator which now extends from OC
up to the end high reflector HR$_2$.

Although there is some interest to examine Josephson coupled BECs
in optical cavities \cite{zhang2008}, here we assume the BECs are
spatially disconnected, such that the condensates are kept in
traps which are sufficiently far apart from each other to avoid
spatial overlap of condensate wave functions. As long as there is
Josephson coupling then condensate numbers would have dynamics and
our treatment with frozen density profile cannot be applied.In
addition to Josephson effect one could consider dense condensates
with interaction terms and dissipation effects to find equilibrium
density profiles so that our treatment can be extended to such a
case to determine electric susceptibility for calculating
possibility of dispersion compensation and reduction of group
velocity.

The BECs are in the EIT
configuration used in ultraslow light experiments \cite{hau_light_1999,matsko_slow_2001}.
The circulating laser pulse acts as the probe field in the BECs
and external coupling lasers are used to achieve transparency.
BEC1 introduces delay to lower the repetition rate while the BEC2
acts as a dispersion compensator. Additional delay is also produced
by BEC2. Axial sizes of the BECs are assumed to be sufficiently
shorter than the  cavity length while we have assumed that proper focusing optics is employed to keep the spotsize within the extent of the BEC, so that the original mode structure was preserved.
Recent experiments with BECs in optical cavities \cite{baumann2010} are successfully modeled under the assumption
of frozen spatial mode profile of the cavity.

Let us assume that the repetition rate of the short
cavity is $f_0=c/2L_0$, where $L_0$ is the effective optical path
length of the short resonator. In the simulations, we assumed
that, without the intracavity BECs, $f_0=25$ MHz which is
easily achievable in standard mode-locked laser configurations. The total
round-trip group delay $T_g$ (in other words, the reverse of the
reduced pulse repetition rate) will be given by
\begin{eqnarray}
T_g  =2\left(\frac{L_1}{v_{g1}} + \frac{L_2}{v_{g2}}\right).
\end{eqnarray}
 Here, we neglected the round-trip
group delay of the initial short cavity, and denoted the effective
length and the group velocity for the condensates ($i=1,2$) as
$L_i$ and $v_{gi}$, respectively, such that \cite{tarhan_dispersive_2006},
\begin{eqnarray}
L_i  = \left[ {\frac{{4\pi }}{N}\int\limits_0^\infty
{rdr\int\limits_0^\infty {z^2 \rho _i (r,z)dz} } } \right]^{1/2}
\end{eqnarray}
and
\begin{eqnarray}
\frac{1}{{v_{gi} }} &=& \frac{1}{c} + \frac{\pi }{\lambda
}\frac{{\partial \chi _i }} {{\partial \Delta }}.
\end{eqnarray}
The effective length is the rms width of the density distribution of the
condensate along the cavity axis ($z$). Here $r$ is the radial coordinate.
In the ideal case, where the inclusion
of the BECs add no loss to the resonator, the average output power
of the laser remains nearly equal to that of the short cavity.
During mode-locked operation of the extended cavity, the
amplification factor A for the pulse energy will therefore be
given by $A = T_g f_0.$
\section{Results}\label{sec:results}
We consider two BECs of $^{23}$Na atoms with parameters
\cite{hau_light_1999}, $M=23$ amu, $a_s = 2.75$ nm, $N_1=N_2=8.3\times10^6$, $\omega_{r}=2\pi\times69$ Hz,
$\omega_{z}=2\pi\times21$ Hz, $\Gamma_3=0.5\gamma$, $\gamma=2\pi\times 10.01$
MHz, and $\Gamma_2=2\pi\times 10^3$ Hz. Resonance wavelength for the probe laser
transition is $\lambda=589$ nm. We take
$\Omega_c=2.5\gamma$ and $\Delta_1=0.01\gamma$.
\subsection{Dispersion compensation}
\begin{figure}
\centering{\vspace{0.5cm}}
\includegraphics[width=3.25in]{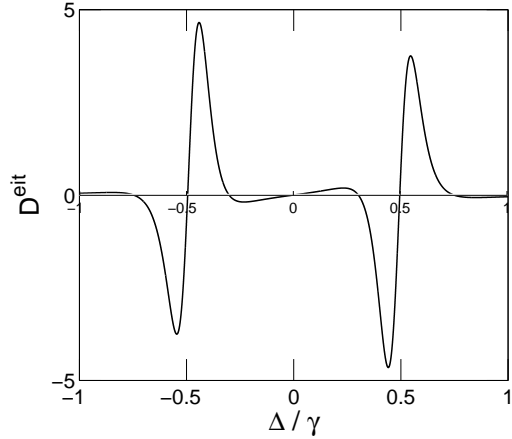}
\caption{Dispersion coefficient for EIT in a BEC of
$^{23}$Na atoms, at temperature $T=381$ nK, with parameters
\cite{hau_light_1999}, $M=23$ amu, $a_s = 2.75$ nm, $N_1=N_2=8.3\times10^6$, $\omega_{r}=2\pi\times69$ Hz,
$\omega_{z}=2\pi\times21$ Hz, $\Gamma_3=0.5\gamma$, $\gamma=2\pi\times 10.01$
MHz, and $\Gamma_2=2\pi\times 10^3$ Hz. Resonance wavelength for the probe laser
transition is $\lambda=589$ nm. We take
$\Omega_c=2.5\gamma$.
The peak density is $\rho_{peak}=6.87\times10^{19}\,(1/{\mathrm m}^3)$. $D$ is
normalized by $1/10^{-10}\,({\mathrm sec}^2)$. Though it is not
visible in the figure, $D(0)= 2.7\times10^{-20}\,({\mathrm
sec}^2)$. At the operating point of the amplifier, $\Delta= -0.01\gamma$, we find
$D =7.9\times10^{-13}\,{\mathrm sec}^2$.}
\label{fig:2}
\end{figure}

Within the transparency window of EIT, imaginary part of the
susceptibility ($\chi^{\prime\prime}$) is very small compared to
the real part of the susceptibility ($\chi^{\prime}$). The group velocity dispersion
$D_i$ is given by $D_i=d^2\phi_i/d\omega^2$ \cite{sennaroglu_photonics_2010},
where $\phi_i=n_i \omega L_i/c$ is the accumulated phase,
with $n_i$ being the refractive index. Using $n_i=\sqrt{1+\chi_i}\approx 1+\chi_i/2$,
$D_i$ can be expressed as
\begin{eqnarray}
D_i = \frac{L_i}{c}\left(\frac{d\chi_i}{d\omega}+\frac{\omega}{2}
\frac{d^2\chi_i}{d\omega^2}\right).
\end{eqnarray}
In the proposed scheme, the second order dispersion of the BEC1 is
positive for a probe pulse slightly blue detuned from the
resonance, which can be seen in Fig. \ref{fig:2}.

At the same but red detuning, dispersion would be almost the same,
but with the opposite sign. For propagation through a single BEC,
this sign change is not relevant on pulse broadening. On the other
hand, by introducing a second BEC, plays a role through the
pulse chirp. The pulses emerging from BEC1 are broadened and
chirped.  In other words, the instantaneous carrier frequency
differs across the temporal pulse profile from the central carrier
frequency.  In the presence of second-order group velocity
dispersion, this chirp is approximately linear near the pulse
center.  If the chirped pulse then enters the second BEC whose
group velocity dispersion parameter is adjusted to be equal in
magnitude but opposite in sign to that of BEC1, the linear chirp
near the pulse center will be cancelled and the initial pulsewidth
will be restored (see Refs. \cite{sennaroglu_photonics_2010,sennaroglu_broadly_2002} and references therein).
For that aim, we assume that BEC2 is also in an EIT configuration but
energy levels for the probe transition are shifted by an external
magnetic field. Hence the blue detuned probe for BEC1 would be
red detuned for BEC2 by the same amount. As the dispersions are of
the same magnitude but of opposite sign, this can be used to
compensate for the positive dispersion of the delay segment
(BEC1), provided that the dispersion introduced in the other intracavity
elements are negligible.

To determine the operating point of the amplifier, we
adjust the parameters of the BECs so that $ D^{BEC1}(x_1)  +
D^{BEC2}(x_2) = 0 $ where $x_1=\Delta_1/\gamma,
x_2=\Delta_2/\gamma$.
The operating point would be independent
of temperature for identical condensates as $D_i$ are proportional to the $\rho_iL_i$ that are the same for
both BECs. Dependence of $D$ on $\Delta$ is shown in Fig. \ref{fig:2} at the chosen temperature of $T=381$ nK.
We find that dispersion compensation
condition, $D_1=-D_2$, is satisfied by $x_1 = -x_2$
at  $\Delta_1/\gamma= 0.01$ which will be the operating point of the amplifier.

We assumed identical traps and condensates only for the sake for simplicity. The main formalism and the dispersion compensation condition $D_1 = - D_2$ is generally valid for non-identical traps and condensates. The effect of different number of atoms in the BECs would be on the operating point (detuning of the probe field) where the dispersion compensation is achieved. We can make a simple estimate. Dispersion coefficient is proportional to the product of BEC density and length, both of which are related to the chemical potential by power laws. Thus the ratio $D_1/D_2$ is related to the ratio of the chemical potentials of the BECs. Assuming Thomas-Fermi profiles, $D_1/D_2$ has negligibly weak temperature dependence, but strong dependence on the ratio of condensed particle numbers $N_1/N_2$. If we know $N_1$ and $N_2$ then the operating point can be determined by simple calculation. Due to number fluctuations the compensation of the dispersion would be incomplete. However, the amplification factors are modest in our case and the pulse duration is in the order of microseconds. For such pulses the dispersion is not too strong and even partial compensation of dispersion should be sufficient and thus we can say fluctuation of particle number would not have crucial effects on our conclusions.
\subsection{Heating rate}

The heating rate of a gaseous sample due to optical absorption is
a standard problem in laser cooling theory (see e.g.
\cite{pethick_bose-einstein_2001} (chapter 4)). The rate of change
in the average kinetic energy of an atom due to absorption can be
evaluated from
\begin{eqnarray}
\frac{1}{2m}\frac{d\bar{p}^2}{dt}=vF_{rad},
\end{eqnarray}
where $v=\hbar k_L/m$ is the recoil velocity of the atom with mass $m$ due to absorption
of a laser photon at wave number $k_L$, and $F_{rad}$ is the radiation force acting on the atom.
Impulse-momentum theorem can be written for the absorption as $F_{rad}\tau_{rad}=\hbar k_L$,
where $\tau_{rad}$ is the characteristic time of interaction between the atom and the radiation field.
Introducing $\Gamma_{rad}=1/\tau_{rad}$, we rewrite the relation as $F_{rad}=\hbar k_L \Gamma_{rad}$.
Second-order perturbation theory can be used to find $\Gamma_{rad}$. Analogous to ac Stark shifts,
we write the energy level shift as $\Delta E=-\frac{1}{2}\alpha\langle {\cal E}^2\rangle$ where $\alpha$
is the single-atom complex polarizability and $\cal{E}$ is the laser pulse amplitude. Imaginary part of the level
shift can be identified by $\Gamma_{rad}=-(2/\hbar)Im(\Delta E)=\alpha^{''}\langle  {\cal E}^2\rangle/\hbar$.
This yields
\begin{eqnarray}
F_{rad}=\frac{1}{\hbar}(\hbar k_L)\alpha^{''}\langle  {\cal E}^2\rangle.
\end{eqnarray}
In our case, $\alpha=\mu_{31}^2L(\Delta)/\hbar$ is the single atom EIT susceptibility, where we introduced a notation
\begin{eqnarray}
L(\Delta)=\frac{{i(-i\Delta  + \Gamma _2 /2)}}{{{(\Gamma _2 /2 -
i\Delta ) (\Gamma _3 /2 - i\Delta ) + \Omega _C ^2 /4}}}.
\end{eqnarray}
The average rate of change of kinetic energy due to absorption
then becomes
\begin{eqnarray}
\frac{1}{2m}\frac{d\bar{p}^2}{dt}=\frac{2}{2m}(\hbar k_L)^2
\frac{\mu_{31}^2\langle  {\cal E}^2\rangle}{\hbar^2}L^{\prime\prime}(\Delta),
\end{eqnarray}
where $L^{\prime\prime}(\Delta)$ is the imaginary part of  $L(\Delta)$.
Identifying the probe Rabi frequency as $\Omega_p^2 = \mu_{31}^2\langle {\cal E}^2\rangle/\hbar^2$,
and by using the equipartition theorem, $k_BT/2=\bar{p}^2/2m$, we finally get the heating rate $\kappa$ to be
\begin{eqnarray}
\kappa=\frac{dT}{dt}=\frac{4}{mk_B}(\hbar k_L)^2\Omega_p^2L^{\prime\prime}(\Delta).
\end{eqnarray}
An additional factor of $2$ is introduced to take into account subsequent emission and absorption processes
together. For the parameters we use in our simulations, taking $\Omega_p=0.1\Omega_c$, the heating rate is evaluated to be $\kappa\sim 1.6$ mK/sec.
We do not assume zero temperature BECs. The heating rate $\kappa$, or the rate of loss of condensed particles, exhibit linear dependence with the temperature and thus one can always choose the initial time ($t_0$) corresponding to the initial temperature ($T_0$) by $t_0 = T_0/\kappa$ so that $T = \kappa t$. We report our results in the figures as functions of temperature which can be translated to time dependence by this scaling transformation.
The effect of dynamically changing temperature $T=\kappa t$ is to make the density of the cloud time dependent.
We shall use semi-ideal model of the BEC \cite{naraschewski_analytical_1998} for an analytical expression of the density of the atomic cloud.

Indeed the main dissipation channel is the loss of the condensate atoms.
In addition to heating, other mechanisms of condensed particle loss may occur, such as three-body loss \cite{kim2004}. Here we do not take them into account as we assume BECs are sufficiently dilute, and the number of particles are not too large. Thus in our case the only source of particle loss out of the condensate is heating, or the recoil momentum transfer, by pulse absorption.
Atoms with sufficient recoil are thermalized and removed from the condensate phase. This in turn reduces the density of the condensate. Once the density becomes lower than the critical density required to maintain the condensate, then the condensate is destroyed and the gas is entirely thermalized. The semi-classical model we employ is in fact is to treat such a case. It contains both the thermal and condensate phases. Below the critical temperature thermal phase serves as the thermal background for the dominant condensate phase. Electromagnetically induced transparency has normally negligible absorption but still it is not zero. Single pulse would transfer a little energy to the condensate. Duration of the interaction due to pulse delay is less than the timeneeded to transfer sufficient energy to destroy the condensate. The recoil energy or the received kinetic energy is translated in our treatment to the temperature variable. Temperature and time has linear relation. The loss of the atoms out of the condensate phase or the condensate fraction reduces by the temperature. These arguments can be put into mathematical context using the semi-ideal model as follows.

The total density at a temperature $T(t)$ is then written to be
\begin {equation}
\label{positiondependentro} \rho(\vec{r},t) =
\frac{\mu(t)-V(\vec{r})}{U_{0}} \Theta(\mu(t)-V(\vec{r}))
\Theta(T_C-T(t)) + \frac{g_{3/2} (z e^{-\beta
V})}{\lambda_T(t)^3},
\end {equation}
where $U_0=4\pi\hbar^2 a_{s}/m$; $m$ is atomic mass; $a_s$ is the
atomic s-wave scattering length; $\mu$ is the chemical potential;
$\Theta(.)$ is the Heaviside step function;
$g_{n}(x)=\Sigma_{j}\,x^j/j^n$ is the Bose function; $\lambda_T$
is the thermal de Br\"{o}glie wavelength; $\beta=1/k_{B}T$;
$z=\exp{(\beta\mu)}$ is the fugacity, and $T_C$ is the critical
temperature which is $T_C=424$ nK in out case. The optical pulse would
heat the cloud to $T_C$ in $265\,\mu$s. The maximum pulse delay time $\sim 65\,\mu$s  is less than the the critical
time at which the condensate turns into thermal gas. However, as the multiple passes of the pulse train over the
condensate continue to heat it we shall consider time, and corresponding temperature range, in our examinations beyond
the critical time and temperature. The external trapping potential is $V(\vec{r})=(m/2)
(\omega_r^2 r^2+\omega_z^2 z^2)$ with $\omega_r$ the radial trap
frequency and $\omega_z$ the angular frequency in the z direction.
$\mu$ is determined from
$N=\int\,\mathrm{d}^3\vec{r}\rho(\vec{r})$. At temperatures below
$T_c$ this yields\cite{naraschewski_analytical_1998}
\begin{eqnarray}
\mu(t)=\mu_{TF}\left(\frac{N_0}{N}(t)\right)^{2/5},
\end{eqnarray}
where $\mu_{TF}$ is the chemical potential evaluated under
Thomas-Fermi approximation and the condensate fraction is given by
\begin{eqnarray}
\frac{N_0}{N}(t)=1-x(t)^3-s\frac{\zeta(2)}{\zeta(3)}x(t)^2(1-x(t)^3)^{2/5},
\end{eqnarray}
with $x(t)=T(t)/T_c$, and $\zeta$ is the Riemann-Zeta function. Thomas-Fermi
profile is valid for small healing length, $\xi=1/\sqrt{8\pi a_s \rho}$, of the condensate relative
to the harmonic trap length.
The scaling parameter $s$, characterizing the strength of atomic
interactions within the condensate, is calculated to
be\cite{naraschewski_analytical_1998}
\begin{eqnarray}
s=\frac{\mu_{TF}}{k_BT_C}=\frac{1}{2}\zeta(3)^{1/3}
\left(15N^{1/6}\frac{a_s}{a_h}\right)^{2/5}.
\end{eqnarray}
Here, $a_h=\sqrt{\hbar/m(\omega_z\omega_r^2)^{1/3}}$ denotes the
average harmonic oscillator length scale. At temperatures above $T_C$, $\mu$ can be determined from
$\mathrm{Li}_3(z)=\zeta(3)/x^3$, where $Li_3(.)$ is the
third-order polylogarithm function. The semi-ideal model has a
wide-range of validity in representing density distribution of a
trapped Bose gas at finite temperature provided that
$s<0.4$ \cite{naraschewski_analytical_1998}. At the same time, the interactions are
assumed to be strong enough to ensure $\mu\gg \hbar\omega_{r,z}$,
so that the kinetic energy of the condensate can be neglected
according to the Thomas-Fermi approximation. In typical slow-light
experiments in cold atomic gases, $s$ remains within these limits.
The time dependence of the density is translated to the condensate expansion
and the group velocity increase which in turn affects the amplification factor
as shown in the following section. We note that decoupling of the matter
wave dynamics from the optical pulse dynamics is based upon their significantly
different time scales. The dynamics of condesate matter wave happens in the $ms$ scale,
while optical pulse evolves in $\mu$s \cite{dutton2004}.  We also ignore higher dimensional
propagation effects in the optical pulse dynamics such as multimode
waveguiding \cite{tarhan2007} and diffraction losses \cite{mustecap2001}.
\subsection{Amplification factor}

\begin{figure}
\centering{\vspace{0.5cm}}
\includegraphics[width=3.25in]{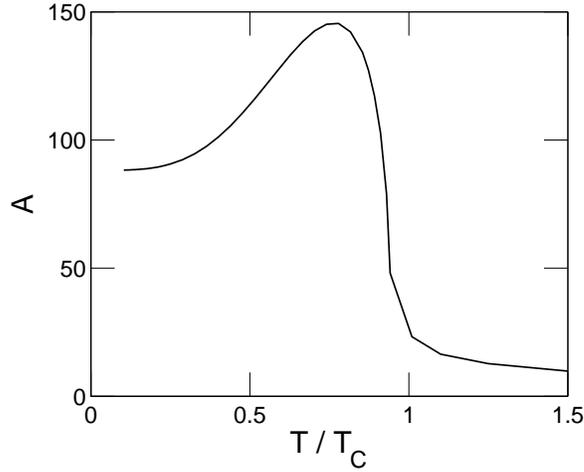}
\caption{Dependence of amplification factor $A$ on temperature, for the same parameters as
in Fig. \ref{fig:2}. The temperature is scaled by the critical temperature $T_C=424$ nK. The heating
rate $\kappa=1.6$ mK/sec translates the same behavior into the time domain via $T=\kappa t$.}
\label{fig:3}
\end{figure}

The result of the temperature dependent amplification calculations is shown in
Fig. \ref{fig:3}. Amplification factor
increases with temperature up to a temperature $T\sim 318$ nK, close but less than the critical temperature
$T_C=424$ nK of the BEC. In the semi-ideal picture of the condensate density profile, this is due to the contribution
from the increasing effect of the thermal component in the pulse propagation via the group velocity and the length
of the condensate.

\begin{figure}[htb]
\centering
\subfigure[]{
\includegraphics[width=7cm]{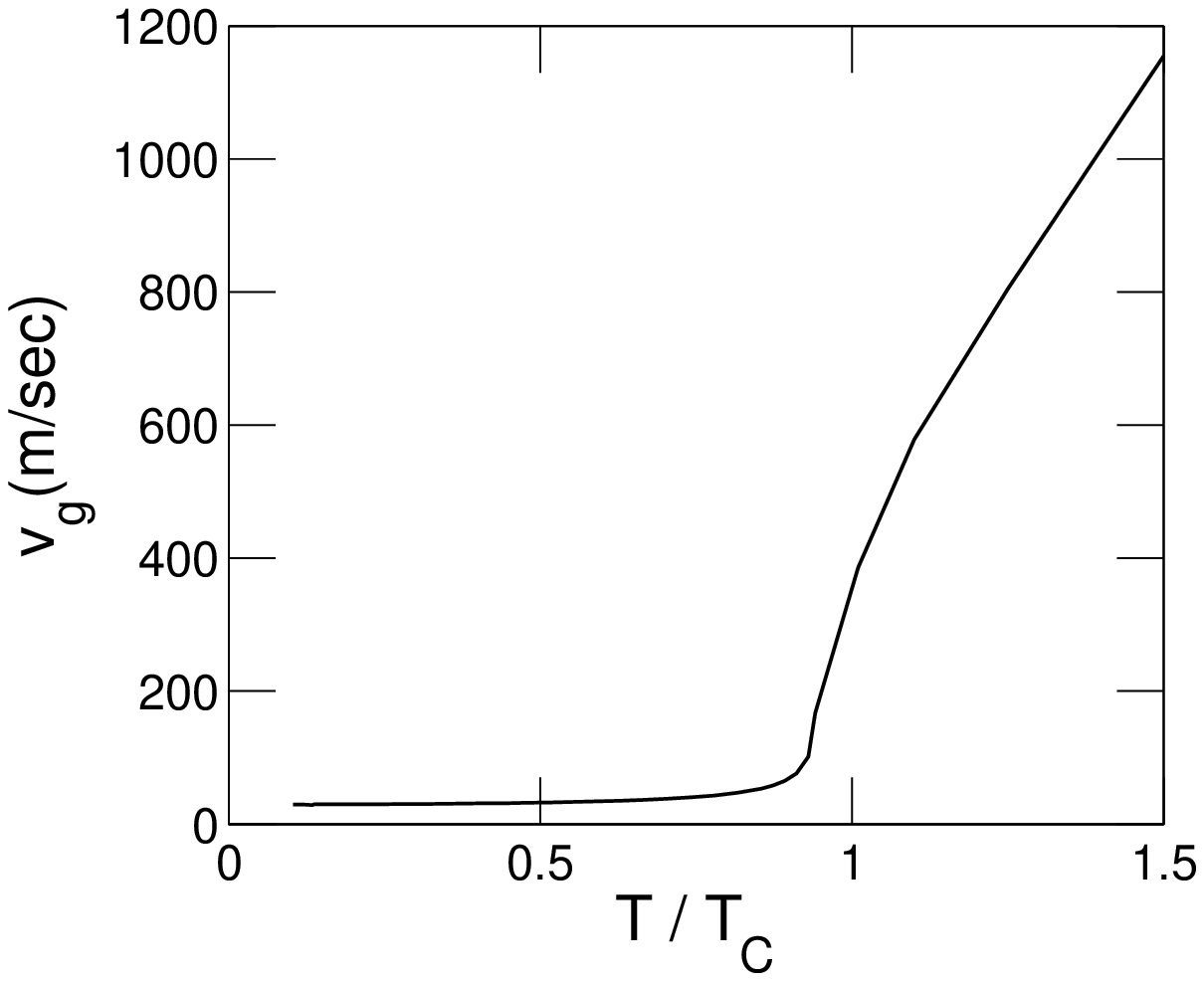}
\label{fig:4a}
}
\subfigure[]{
\includegraphics[width=7cm]{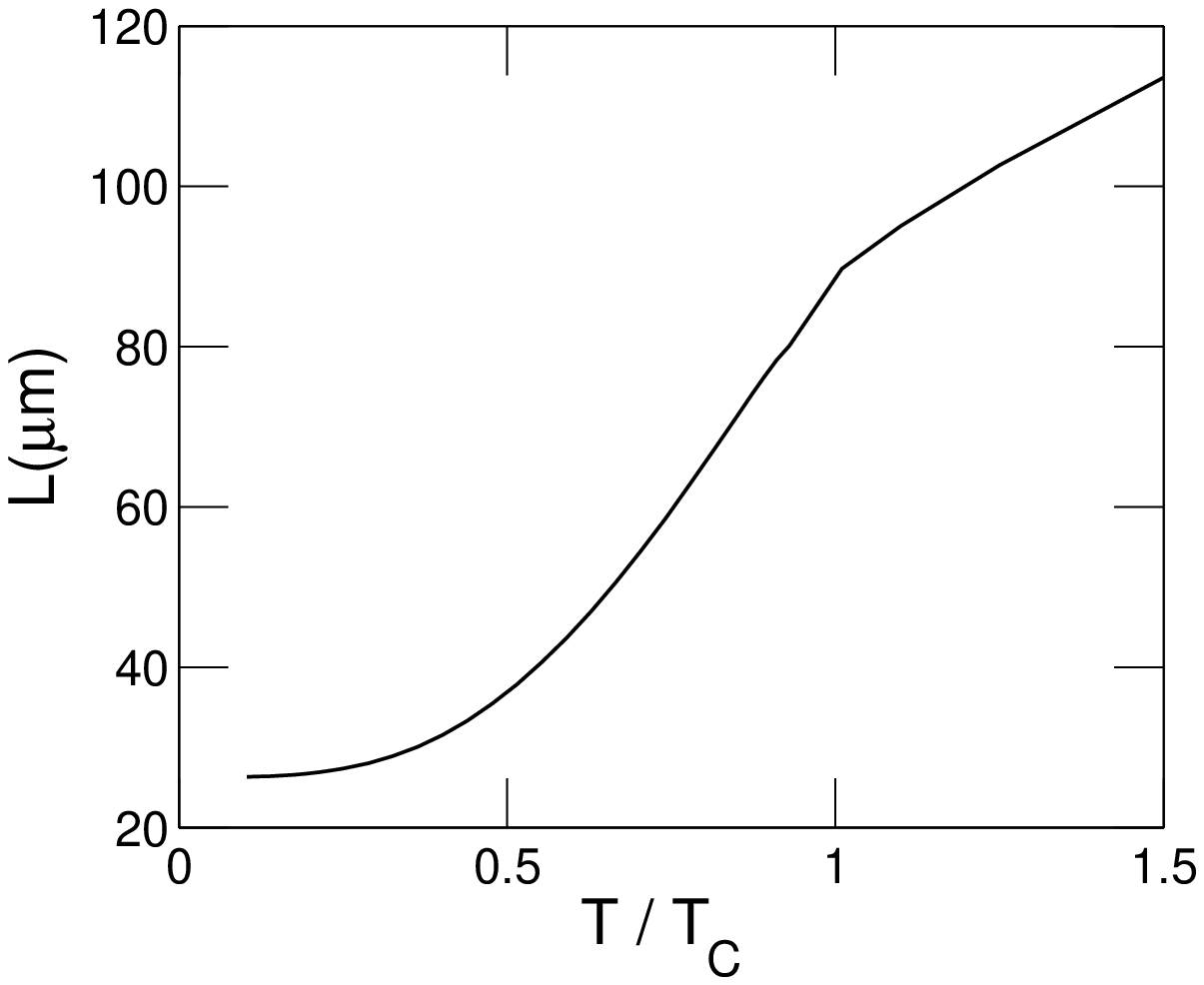}
\label{fig:4b}
}
\caption{(a) Thermal behavior of the group velocity of the probe pulse. (b) Thermal
behavior of the effective length of the BEC. The parameters used for both curves
are the same as
in Fig. \ref{fig:2}. The temperature is scaled by the critical temperature $T_C=424$ nK. The heating
rate $\kappa=1.6$ mK/sec translates the same behaviors into the time domain via $T=\kappa t$.
}
\label{fig:4}
\end{figure}

As can be seen in Fig. \ref{fig:4}, the group velocity weakly depends on the temperature
in the condensate phase. Thermal behavior of the group velocity can be explained by examining the thermal
behavior of the condensate density. In the semi-ideal model that we use both the condensate
and the thermal gas components are considered. Below the critical temperature,
the condensate component emerges sharply and dominates over the thermal background gas.
Density varies little at low temperatures. Thus the ultraslow propagation speeds weakly change
with the temperature. After the critical temperature, there is no more condensate and the
thermal gas gets more rapidly diluted with the temperature, causing the kink in Fig.  \ref{fig:4a},
which is also observed experimentally \cite{hau_light_1999}.

The main effect of the temperature is the expansion of the
condensate which causes the increase of the delay times of the pulse through the condensate and
hence $T_g$ increases. Behavior of the amplification factor with the temperature
follows that of the effective length. Beyond $T_c$, almost linear
behavior of group velocity and the effective length with temperature result in
weak dependence of $T_g$ and hence amplification on the temperature. The amplification is
larger in the condensation regime due to large group delays. Fig. \ref{fig:4a} shows that
group velocity of the pulse in the thermal gas regime rises sharply and it is
faster than expansion of the atomic cloud shown in Fig. \ref{fig:4b}.
Accordingly, amplification factor sharply drops after a critical temperature close to $T_c$ and continues
to decrease slowly in the thermal gas phase.

\subsection{Spectral bandwidth}

Finally, we investigate the spectral bandwidth limitations of the
BEC. Ideally, it is desirable to have a system which has a very
broad transparency to support the propagation of pulses with very
short duration.  To provide a feel for how short a pulse the BECs
can support, we calculate the spectral bandwidth $\Delta \nu$
that corresponds to the transparency window of the BEC in the EIT scheme.
The net bandwidth due to both of the BECs is $2\Delta \nu$
so that the temporal width  $\tau_p$ of the pulse that can be
supported can be estimated by $\tau _p  = 1/2\Delta \nu$.
Using \cite{fleischhauer_electromagnetically_2005}
\begin{eqnarray}
\Delta\nu = \frac{\Omega_c^2}{\gamma}\frac{1}{\sqrt{\kappa_{\nu}}},
\end{eqnarray}
with $\kappa_{\nu}=3\rho\lambda^3(k_LL)/8\pi^2$ being the opacity of the atomic cloud of length
$L$ and density $\rho$, we find $\Delta\nu\sim 0.1\gamma$. Note that the operation
point $\Delta=0.01\gamma$ lies within the transparency window. The pulses that
can be supported by the condensates should have widths of the order of $\sim \mu$s.
Pulses of shorter widths could be supported by considering larger $\Omega_c$. The
cost however would be to get lower amplification factors as the group velocity would
increase with increasing Rabi frequency of the control field.  More ingenious designs
that specifically considers transparency window enhancement for broadband pulses
are proposed \cite{yavuz2007} but their integration to the present proposal require further
studies.
\section{Conclusions}\label{sec:conclusion}

In conclusion,  we have investigated the feasibility of using
Bose-Einstein condensates for laser pulse amplification.  The
method involves the introduction of two BECs inside the resonator
of a passively mode-locked laser.  The large delay produced by the
BECs lowers the pulse repetition rate and scales up the output
energy. Our calculations show that pulse amplification factors of
the order $\sim 10^2$ should be possible over a condensate length
of $\sim 50\,\mu$m. We further showed that a second BEC could be
used to provide dispersion compensation. However, the
amplification factor decreases with time due to the presence of
optical absorption. We have estimated the heating rate to be about
$1.6$ mK/sec, which severly limits the operation time of the
system in the condensate regime. A critical time of operation for
optimum amplification is found to be about $200\,\mu$s. That would
require an additional optical switching to extract the pulse at
the right time out of the cavity. Alternatively, reducing the
effects of absorption by considering multi-level EIT systems
\cite{hemmer_efficient_1995,wang_electromagnetically_2004} or
designing a compensating simultaneous cooling mechanism on the
atomic cloud in EIT scheme
\cite{schmidt-kaler_laser_2001,evers_double-eit_2004} can be
considered. This would eliminate the need of reconstruction of the
BEC to amplify different pulses at optimum conditions. For quick
and easy generation of BECs atom chips can be promising
\cite{hansel2001}. To aim at higher intensities and amplification
factors, the nonlinear response of the atomic gas should also be
taken into account.  Denser condensates, with nonlinear response
and quantum corrections, including atom-atom interactions, but
without local field correction, seem to be beneficial for larger
delay times as well \cite{haghshenasfard_controlling_2012}.

\ack

D.T. acknowledges the support from TUBITAK (The Scientific and Technological Research
Council of Turkey) Career Grant No 109T686. \"O. E. M.
is supported by TUBITAK under project TBAG-109T267. D. T. thanks G. S. Agarwal
for helpful discussions.
\section*{References}



\end{document}